\documentclass[aps,preprint,nofootinbib,superscriptaddress,prc,floatfix]{revtex4-1}
\usepackage{amssymb}
\usepackage{amsmath}
\usepackage{epsfig}
\usepackage{epstopdf}
\usepackage[bookmarksnumbered,bookmarksopen,colorlinks,citecolor=blue,linkcolor=blue]{hyperref}
\usepackage{booktabs}
\usepackage{array}
\usepackage{longtable}
\usepackage{bm}
\begin{document}

\date{\today}
\title{Nuclear mass calculation based on anisotropic kernel ridge regression and $\alpha$-decay chain recursion}

\author{Jiaxin Gui}\thanks{These authors contributed equally to this work.}
\affiliation{Department of Physics, Guangxi Normal University, Guilin 541004, People's Republic of China}
\author{Minghui Hu}\thanks{These authors contributed equally to this work.}
\affiliation{Department of Physics, Guangxi Normal University, Guilin 541004, People's Republic of China}
\author{Xuan Zhang}
\affiliation{Department of Physics, Guangxi Normal University, Guilin 541004, People's Republic of China}
\author{Pengfei Ma}
\affiliation{Department of Physics, Guangxi Normal University, Guilin 541004, People's Republic of China}
\author{Junlong Tian}
\email{tianjl@gxnu.edu.cn}
\affiliation{Department of Physics, Guangxi Normal University, Guilin 541004, People's Republic of China}
\affiliation{Guangxi Key Laboratory of Nuclear Physics and Technology, Guilin 541004, People's Republic of China}
\author{Xinrui Zhang}
\email{zhangxinrui@gxnu.edu.cn}
\affiliation{Department of Physics, Guangxi Normal University, Guilin 541004, People's Republic of China}
\affiliation{Guangxi Key Laboratory of Nuclear Physics and Technology, Guilin 541004, People's Republic of China}

\begin{abstract}
Many short-lived or rarely produced nuclei defy direct mass spectrometry, but
their masses are constrained by the $\alpha$-decay energies connecting them
through decay chains. We combine anisotropic kernel ridge regression (AKRR),
an extended Garvey--Kelson relation (GK16), and $\alpha$-decay chain
recursion: AKRR learns the residuals between experimental $\alpha$-decay
energies and WS4 predictions, GK16 removes residual local fluctuations, and
the corrected energies convert a measured anchor mass into the target mass by
exact recursion---replacing absolute-mass extrapolation with interpolation of
relative decay energies plus exact algebra. The only data-adjusted quantities
are the three kernel hyperparameters, fixed by leave-one-out cross
validation. For the 22 nuclei whose masses were measured only after AME2003, the $Q_\alpha$ RMSD
falls from 169~keV (WS4) to 108~keV (AKRR) and 65~keV (AKRR+GK16), and the
mass-excess RMSD reaches 86~keV, versus 348 and 211~keV for WS4 and AKRR. Mass excesses are
further predicted for the 122 nuclei with measured $Q_\alpha$ but no directly
measured masses.
\end{abstract}

\maketitle

\section{Introduction}
\label{sec:intro}
Nuclear masses are needed across nuclear structure, nuclear astrophysics,
and tests of fundamental symmetries, and they are a key input to modeling of
$r$-process nucleosynthesis~\cite{Wallace81,Wormer94,Niuzm22,Schatz98} and
neutron-star crusts. For superheavy nuclei (SHN), the mass sets the $Q$
values of decay chains and the competition between $\alpha$ decay and
spontaneous fission, and thus the location of the predicted ``island of
stability''. Direct mass spectrometry of SHN, however, is severely limited by
their extremely low production cross sections and short half-lives: even for
the well-studied $^{294}$Og, reported in Ref.~\cite{Oganes06}, only a sequence
of $\alpha$-decay energies is available. Nor is this an isolated case:
AME2020 contains 122 nuclei with measured $\alpha$-decay energies
($Q_{\alpha}$) but no direct mass determination (Table~\ref{tab:predictions}).
Since $Q_{\alpha}$ constrains the mass difference between parent and
daughter, these measurements pin down the unknown masses indirectly but
tightly, provided the decay energies can be converted into mass excesses
without errors accumulating along the chain.

Theoretical mass predictions follow three main strategies. Global
models---the Weizs\"acker--Skyrme (WS) model~\cite{Ning10,liumin11,wangn11},
FRDM12~\cite{Moller16}, the Duflo--Zuker model~\cite{DZ95}, and the
Skyrme--Hartree--Fock--Bogoliubov approach~\cite{Gori01,Gori10,Geng05}---reproduce
the 2340 known masses with root-mean-square deviations (RMSDs) of
0.3--0.7~MeV, still far above the $\sim$100~keV required for $r$-process
simulations~\cite{Mump15}, and their extrapolation uncertainties can reach
tens of MeV. Local mass relations express a mass through its neighbors; they
are accurate near measured nuclei but degrade rapidly away from them, as in
the Audi--Wapstra systematics~\cite{Audi93,Audi95,Audi03}, Garvey--Kelson
relations~\cite{Garv66,Janec74,Mona77,Janec85,Bare05,Bare08,Hirs08,Mora11},
and residual proton--neutron interaction
relations~\cite{Fugj10,Jiangh10,Fugj11,Jiangh12}. Machine learning offers a
third approach: radial basis function
networks~\cite{wangn11,Niuzm18}, Bayesian neural
networks~\cite{Niuzm22}, and kernel ridge regression
(KRR)~\cite{Wuxh20,Wuxh23} with its anisotropic variant
(AKRR)~\cite{Wuxh24,Tian25} considerably improve the description of known
masses, though reliable extrapolation beyond the measured frontier remains
open.

In this work we operate on $Q_\alpha$ rather than directly on mass excesses:
given one known mass along a connected decay chain, all remaining masses
follow from the decay energies. This suits heavy and superheavy nuclei, where
decay measurements are more accessible than direct mass determinations. WS4
$Q_\alpha$ values are first corrected by AKRR trained on measured residuals,
refined by the extended Garvey--Kelson relation GK16, and converted into mass
excesses by backward recursion from a measured anchor nucleus. Every step
acts on the same observable, each component compensating the previous one:
AKRR removes global model bias, GK16 damps the residual local fluctuations,
and the recursion maps accurate $Q_\alpha$ inputs to masses. The motivation
is an intrinsic asymmetry between the two observables: a mass excess is
absolute, so predicting it outside the measured region is an extrapolation,
whereas an $\alpha$-decay energy is a relative quantity whose estimation from
measured neighbors is an interpolation, controllable at the tens-of-keV
level. The recursion itself is exact algebra, so the only model error
reaching the final mass is the interpolation uncertainty of individual
$Q_\alpha$ values along the chain. Adopting AME2020 as reference, we validate
the scheme on 22 nuclei whose masses were only estimated in AME2003 and
subsequently measured (Sec.~\ref{sec:test}), then apply it to the 122 nuclei
with measured $Q_\alpha$ but no direct mass (Sec.~\ref{sec:mass122}).

\section{Theoretical Framework}
\label{sec:framework}
The calculation proceeds along a fixed chain of transformations,
\begin{equation}
Q_\alpha^{\rm WS4}\ \xrightarrow{\ \rm AKRR\ }\
Q_\alpha^{\rm AKRR}\ \xrightarrow{\ \rm GK16\ }\
Q_\alpha^{\rm AKRR+GK16}\ \xrightarrow{\ \rm MRQ\ }\
M(N,Z),
\label{eq:pipeline}
\end{equation}
where the only quantities adjusted to the data are the three kernel
hyperparameters, fixed by leave-one-out cross validation (LOO CV).
Section~\ref{sec:mrq} sets out the mass recursion (MRQ), which determines
the quantity that the first two arrows must supply; Secs.~\ref{sec:akrr}
and~\ref{sec:gk16} then describe how that $Q_\alpha$ input is built.

\subsection{Mass recursion along the $\alpha$-decay chain (MRQ)}
\label{sec:mrq}

The $\alpha$-decay energy $Q_\alpha(N,Z)$ of a nucleus with $N$ neutrons and
$Z$ protons is expressed through the atomic mass excesses of the parent and
daughter as
\begin{equation}
Q_\alpha(N,Z)=M(N,Z)-M(N-2,Z-2)-M_\alpha,
\label{eq:Q_alpha_conservation}
\end{equation}
where $M_\alpha=2.4249$~MeV is the atomic mass excess of
$^{4}$He~\cite{AME2020}. All quantities being atomic mass excesses,
the electron rest masses cancel, and electron-binding effects are already included in the atomic masses.
Throughout this work $Q_\alpha$ refers to
ground-state-to-ground-state transitions, i.e., directly to the
parent--daughter mass difference; decays populating excited daughter states
are excluded. The relation is independent of the spin and parity of the
ground states and holds for even-even, odd-$A$, and odd-odd nuclei alike,
whether the value is experimental or predicted.

For $n$ successive $\alpha$ decays along a single decay chain,
\begin{align}
Q_{\alpha 1} &= M(N,Z)-M(N-2,Z-2)-M_\alpha, \label{eq:Qa1}\\
Q_{\alpha 2} &= M(N-2,Z-2)-M(N-4,Z-4)-M_\alpha, \label{eq:Qa2}\\
Q_{\alpha 3} &= M(N-4,Z-4)-M(N-6,Z-6)-M_\alpha, \label{eq:Qa3}\\
&\quad\cdots \nonumber\\
Q_{\alpha n} &= M(N-2n+2,Z-2n+2)-M(N-2n,Z-2n)-M_\alpha . \label{eq:Qan}
\end{align}
Summing Eqs.~\eqref{eq:Qa1}--\eqref{eq:Qan} cancels all intermediate mass
excesses: each of them appears once with a plus sign and once with a minus
sign, so only the two ends of the chain survive, and the mass excess of the
parent nucleus follows as
\begin{equation}
M(N,Z)=M(N-2n,Z-2n)+\sum_{i=1}^{n}Q_{\alpha i}+nM_\alpha ,
\label{eq:mass_recursion}
\end{equation}
where $(N-2n,Z-2n)$ is the anchor nucleus. The chain is traced from the
target down to this known anchor, and the mass is reconstructed backward from
the anchor to the target.

Equation~\eqref{eq:mass_recursion} is exact and underlies the recursion
scheme. Three properties deserve emphasis. First, only mass differences
appear, so any overall offset of the underlying mass model cancels and the
final accuracy is governed solely by the individual $Q_\alpha$ values.
Second, the anchor-mass uncertainty propagates fully to the target while the
$Q_\alpha$ uncertainties accumulate along the chain; the anchor should
therefore be chosen as close to the target as the data permit. Third, the
recursion transforms the prediction task: evaluating the absolute target
mass by a global model is an extrapolation, whereas each $Q_\alpha$ in the
sum is a relative quantity linking adjacent nuclei and may be taken from
experiment or from the local AKRR+GK16 predictor, constrained by measured
$Q_\alpha$ residuals of neighboring nuclei. A mass-extrapolation problem
thus becomes a $Q_\alpha$ interpolation followed by exact
algebra.

In practice, we trace the chain from the target toward lower mass number
until a daughter with an experimentally known mass excess is reached;
AME2020~\cite{AME2020} is used for the production calculations and
AME2003~\cite{Audi03} for the benchmark of Sec.~\ref{sec:test}. Experimental
ground-state $Q_\alpha$ is adopted for each link when available; missing
links are filled by AKRR+GK16 predictions built upon the WS4 mass
model~\cite{Wangn14,Wuxh24}. The measured anchor sets the absolute scale,
measured decay energies constrain the mass differences, and theory enters
only where data are absent. The two $Q_\alpha$ providers are described next.
\subsection{Anisotropic kernel ridge regression (AKRR)}
\label{sec:akrr}
AKRR was proposed by X.~H. Wu and C.~Pan~\cite{Wuxh24}, who added to
standard KRR a superposition of two two-dimensional Gaussian kernels,
motivated by the stronger correlations among isotopes and isotones than
among isobars---hence distinct correlation lengths along the neutron and
proton axes. The hyperparameters are fixed by cross validation, and LOO
tests confirm that AKRR improves global mass models. Here we
apply AKRR not to masses but directly to $\alpha$-decay energies, the
observable propagated through Eq.~\eqref{eq:mass_recursion}.

In this scheme, the predicted $\alpha$-decay-energy residual
$S_\alpha(N,Z)$ reads
\begin{equation}
S_{\alpha}(N,Z) = \sum_{j=1}^{m} K\bigl[(N,Z),\,(N_j,Z_j)\bigr]\,\beta_j,
\label{eq1}
\end{equation}
where $m$ is the training-set size, $K$ the kernel linking the target
$(N,Z)$ to the training nuclide $(N_j,Z_j)$, and $\beta_j$ the regression
coefficients, obtained by minimizing
\begin{equation}
L(\boldsymbol{\beta})=\sum_{i=1}^{m}\left[S_{\alpha}(N_i,Z_i)-R_{\alpha}(N_i,Z_i)\right]^{2}+\lambda\|\boldsymbol{\beta}\|^{2},
\label{eq2}
\end{equation}
with the training residual
\begin{equation}
R_\alpha(N,Z) = Q_\alpha^{\mathrm{exp}}(N,Z) - Q_\alpha^{\mathrm{th}}(N,Z),
\label{eq:R_alpha}
\end{equation}
$\lambda$ the regularization strength, and $Q_\alpha^{\mathrm{th}}$ the WS4
value. Minimization yields the closed form
\begin{equation}
\boldsymbol{\beta}=(\boldsymbol{K}+\lambda\boldsymbol{I})^{-1}\boldsymbol{R}_{\alpha},
\label{eq3}
\end{equation}
where $\boldsymbol{K}$ is the $m\times m$ kernel matrix,
$\boldsymbol{I}$ the identity, and $\boldsymbol{R}_\alpha$ the residual
vector. Solving Eq.~\eqref{eq3} and substituting into Eq.~\eqref{eq1} gives
the AKRR-corrected energy
\begin{equation}
Q_\alpha^{\mathrm{AKRR}}(N,Z) = Q_\alpha^{\mathrm{th}}(N,Z) + S_\alpha(N,Z),
\label{eq:Q_AKRR}
\end{equation}
the first arrow of Eq.~\eqref{eq:pipeline}.

Following Ref.~\cite{Wuxh24}, the kernel is the sum of two anisotropic
two-dimensional Gaussians:
\begin{equation}
\begin{split}
&K[(N, Z), (N_j,Z_j)]\\
&=\exp\left\{-\left[\frac{(N - N_j)^2}{2\sigma_1^2}+\frac{(Z - Z_j)^2}{2\sigma_2^2}\right]\right\}\\
&+\exp\left\{-\left[\frac{(N - N_j)^2}{2\sigma_2^2}+\frac{(Z - Z_j)^2}{2\sigma_1^2}\right]\right\},
\end{split}
\label{eq6}
\end{equation}
assigning a common correlation length to the neutron and proton directions
and a shorter one to the two tilted directions of the nuclear chart.
We select 2290 experimental $Q_\alpha$ values (including negative ones) with $Z\geq10$, $N\geq10$ and uncertainties below
100~keV, comprising 2263 entries from the AME2020 database~\cite{AME2020} and 27 additional nuclei from recent publications ($^{160}\mathrm{Os}$ \cite{160Os}, $^{190}\mathrm{At}$ \cite{190At},  $^{203}\mathrm{Ac}$ \cite{203Ac}, $^{204}\mathrm{Ac}$ \cite{204Ac}, $^{207}\mathrm{Th}$ \cite{207Th}, $^{210}\mathrm{Pa}$ \cite{210Pa}, $^{214}\mathrm{U}$ \cite{214216218U},  $^{227}\mathrm{Pu}$ \cite{227Pu},  $^{231}\mathrm{Am}$,  $^{235}\mathrm{Bk}$ \cite{231Am235Bk},  $^{249}\mathrm{No}$, $^{253}\mathrm{Rf}$ \cite{249No253Rf}, $^{251}\mathrm{Lr}$ \cite{251Lr}, $^{257}\mathrm{Sg}$ \cite{257Sg}, $^{265,267}\mathrm{Sg}$, $^{269,271}\mathrm{Hs}$, $^{275}\mathrm{Ds}$ \cite{265-267Sg271Hs275Ds}, $^{272}\mathrm{Hs}$, $^{276}\mathrm{Ds}$ \cite{272Hs276Ds}, $^{284}\mathrm{Cn}$ \cite{284Cn}, $^{286}\mathrm{Mc}$ \cite{286Mc}, $^{277}\mathrm{Ds}$, $^{284}\mathrm{Fl}$, and $^{288-289}\mathrm{Lv}$ \cite{288Lv}). For each hyperparameter candidate, the corrected $Q_\alpha$ of every
nucleus is computed from a model trained on the remaining 2289, and the LOO
RMSD between corrected and experimental values serves as the validation
score. Minimization yields $\lambda=0.3$, $\sigma_1=3.3$, $\sigma_2=0.7$, with an RMSD of $132\,\text{keV}$. These values are consistent with the mass-residual fit reported in Ref.~\cite{Wuxh24}, indicating similar residual structures.

\subsection{Extended Garvey--Kelson relation (GK16) for $Q_{\alpha}$}
\label{sec:gk16}
Masses enter the workflow only as measured anchors and as the final output of
Eq.~\eqref{eq:mass_recursion}; the correction scheme below operates solely on
$Q_\alpha$, and the quantities $M(N,Z)$ are used only for algebraic
derivation.

\emph{(1) The original GK relations.}
Garvey and Kelson proposed a family of accurate local mass
relations~\cite{KG66,KG88} in transverse and longitudinal variants, each
connecting six adjacent nuclei:
\begin{align}
\mathcal{G}_{T} &= M(N-2,Z+2)-M(N,Z)+M(N-1,Z) \nonumber\\
&\quad -M(N-2,Z+1)+M(N,Z+1)-M(N-1,Z+2) \approx 0,
\label{eq:GKT}\\
\mathcal{G}_{L} &= M(N+2,Z-2)-M(N,Z)+M(N+1,Z) \nonumber\\
&\quad -M(N+2,Z-1)+M(N,Z-1)-M(N+1,Z-2) \approx 0.
\label{eq:GKL}
\end{align}
Collecting all constraints containing $M(N,Z)$ and averaging the twelve
resulting estimates yields a relation coupling 21 nuclei, GK12~\cite{Bare08}.
GK12, however, excludes the four diagonal neighbors $(N\pm1,Z\pm1)$---precisely
the nuclei most strongly correlated with $(N,Z)$ in the AKRR kernel---which
motivates the extended GK16 relation used here.

\emph{(2) Applicability of the GK relations to $Q_{\alpha}$.}
Why can mass-based GK relations be transferred to $Q_\alpha$? The GK
operators $\mathcal{G}_{T,L}$ are linear with coefficients summing to zero,
so any additive constant---in particular $M_\alpha$ or a global offset of the
mass scale---is eliminated. Applying $\mathcal{G}_{T,L}$ to
Eq.~\eqref{eq:Q_alpha_conservation} then gives
\begin{equation}
\mathcal{G}_{T,L}\,Q_{\alpha}(N,Z)
=\mathcal{G}_{T,L}\,M(N,Z)-\mathcal{G}_{T,L}\,M(N-2,Z-2)\approx0.
\end{equation}
Each GK relation for $Q_{\alpha}$ is thus the difference of two GK mass
relations centered at $(N,Z)$ and $(N-2,Z-2)$ and retains the same form;
all GK-type relations below apply equally to $Q_{\alpha}$, including
negative values. We write $Q$ for $Q_\alpha$ in the remainder of this
subsection.

\begin{figure}[!htbp]
\includegraphics[angle=0,width=0.99\textwidth]{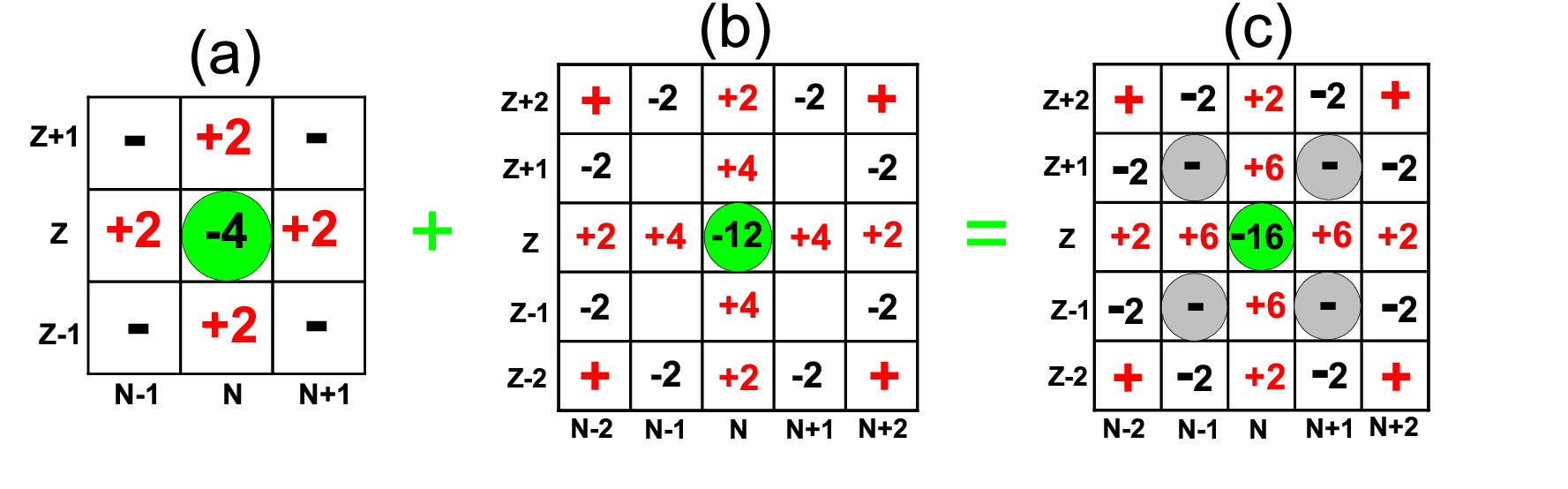}
\caption{(Color online) Schematic construction of the local GK16 relation on the $(N,Z)$
plane: (a) the relation derived from residual proton--neutron
interactions~\cite{Fugj11} (Eq.~(\ref{eq15})), (b) the GK12
relation~\cite{Bare08}, and (c) their sum, defining the GK16 relation
(Eq.~(\ref{eq16})). Each symbol marks a nucleus on the nuclidic chart centered
at \((N,Z)\) (solid green circle). Red plus and black minus signs indicate
whether the corresponding \(Q_\alpha\) values are added or subtracted to
yield a null result. The four diagonal neighbors included in GK16 but absent
from GK12 are marked by solid gray circles.}
\label{fig1}
\end{figure}

\emph{(3) Residual proton--neutron interaction relation.}
To include the diagonal neighbors, we use local relations built on the
residual proton--neutron interaction~\cite{Fugj10,Jiangh10,Fugj11,Jiangh12}.
The empirical valence interaction is the double difference
\begin{equation}
\delta Q_{1n\text{-}1p}(N,Z)=Q(N,Z)+Q(N-1,Z-1)-Q(N-1,Z)-Q(N,Z-1),
\label{eq7}
\end{equation}
with $Q$ denoting $Q_\alpha$. Shifting indices gives three further relations
containing $Q(N,Z)$:
\begin{align}
\delta Q_{1n\text{-}1p}(N+1,Z+1)&=Q(N+1,Z+1)+Q(N,Z)\nonumber\\
&\quad -Q(N,Z+1)-Q(N+1,Z), \label{eq8}\\
\delta Q_{1n\text{-}1p}(N,Z+1)&=Q(N,Z+1)+Q(N-1,Z)\nonumber\\
&\quad -Q(N,Z)-Q(N-1,Z+1), \label{eq9}\\
\delta Q_{1n\text{-}1p}(N+1,Z)&=Q(N+1,Z)+Q(N,Z-1)\nonumber\\
&\quad -Q(N+1,Z-1)-Q(N,Z). \label{eq10}
\end{align}
Solving each for $Q(N,Z)$ yields four independent estimates:
\begin{align}
Q(N,Z)&=Q(N,Z-1)+Q(N-1,Z)-Q(N-1,Z-1)\nonumber\\
&\quad +\delta Q_{1n\text{-}1p}(N,Z), \label{eq11}\\
Q(N,Z)&=Q(N,Z+1)+Q(N+1,Z)-Q(N+1,Z+1)\nonumber\\
&\quad +\delta Q_{1n\text{-}1p}(N+1,Z+1), \label{eq12}\\
Q(N,Z)&=Q(N,Z+1)+Q(N-1,Z)-Q(N-1,Z+1)\nonumber\\
&\quad -\delta Q_{1n\text{-}1p}(N,Z+1), \label{eq13}\\
Q(N,Z)&=Q(N,Z-1)+Q(N+1,Z)-Q(N+1,Z-1)\nonumber\\
&\quad -\delta Q_{1n\text{-}1p}(N+1,Z). \label{eq14}
\end{align}
Since the residual proton--neutron interaction varies smoothly with $N$ and
$Z$, we adopt $\delta Q_{1n\text{-}1p}(N,Z)+\delta
Q_{1n\text{-}1p}(N+1,Z+1) \approx\delta Q_{1n\text{-}1p}(N,Z+1)+\delta
Q_{1n\text{-}1p}(N+1,Z)$. Summing Eqs.~\eqref{eq11}--\eqref{eq14} under this
approximation cancels all $\delta Q_{1n\text{-}1p}$ terms and connects the
central nuclide with its eight nearest neighbors,
\begin{align}
&2[Q(N,Z-1)+Q(N,Z+1)+Q(N-1,Z)+Q(N+1,Z)]\nonumber\\
&\quad -[Q(N-1,Z+1)+Q(N+1,Z-1)+Q(N-1,Z-1)+Q(N+1,Z+1)]\nonumber\\
&\quad -4Q(N,Z)\approx0,
\label{eq15}
\end{align}
which, unlike GK12, contains the four diagonal neighbors explicitly. Note
that $\delta Q_{1n\text{-}1p}$ here is a double difference of
$\alpha$-decay energies of four neighboring nuclei---the sense in which the
nine nuclei of Fig.~\ref{fig1}(a) enter---not the residual $p$--$n$
interaction extracted from masses, so $Q$ may take either sign.

\emph{(4) Construction of GK16.}
The construction rests on two approximations: the negligible $N,Z$ dependence
of the exchange terms intrinsic to the GK relations and GK12, and the
smoothness of the residual proton--neutron interaction invoked when combining
Eqs.~\eqref{eq11}--\eqref{eq14}, which also produces the alternating sign
pattern across successive neighbor shells in Eq.~\eqref{eq16}, visible in
Fig.~\ref{fig1}(c). Both rest on the weak variation of the valence-nucleon
interaction over a few units of $N$ and $Z$, and neither holds where
neighboring nuclei lack measured data, i.e., in the extrapolation region
targeted here. GK16 therefore mainly suppresses residual fluctuations where
the AKRR input is already reliable, and its relative weight diminishes toward
the frontier.

GK12 provides twelve estimates of $Q(N,Z)$ with central coefficient 12, and
Eq.~\eqref{eq15} four more with central coefficient 4; direct summation
assigns unit weight to all sixteen estimates, yielding a central term
$16Q(N,Z)$. Each relation is invariant under a global shift $Q \to Q+\Delta$
because its coefficients sum to zero, and the combined relation inherits
this invariance; the central term stands on the left-hand side merely for
compactness. The resulting extended Garvey--Kelson relation, GK16, reads
\begin{eqnarray}
\nonumber16Q(N,Z)&&\approx Q(N-2,Z+2)+Q(N+2,Z-2)+Q(N-2,Z-2)+Q(N+2,Z+2)\\
\nonumber&&-2\left[Q(N-1,Z-2)+Q(N+1,Z+2)+Q(N+1,Z-2)+Q(N-1,Z+2)\right]\\
\nonumber&&-2\left[Q(N-2,Z+1)+Q(N+2,Z-1)+Q(N-2,Z-1)+Q(N+2,Z+1)\right]\\
\nonumber&&-\left[Q(N-1,Z+1)+Q(N+1,Z-1)+Q(N-1,Z-1)+Q(N+1,Z+1)\right]\\
\nonumber&&+2\left[Q(N-2,Z)+Q(N+2,Z)+Q(N,Z+2)+Q(N,Z-2)\right]\\
&& +6\left[Q(N,Z-1)+Q(N,Z+1)+Q(N-1,Z)+Q(N+1,Z)\right].
\label{eq16}
\end{eqnarray}

The construction is illustrated in Fig.~\ref{fig1}, with panel (a) plus
panel (b) yielding panel (c): GK16 merges sixteen estimates---the twelve of
GK12 plus the four of Eqs.~\eqref{eq11}--\eqref{eq14}---into a single
relation coupling 25 nuclei [Fig.~\ref{fig1}(c)], including the four
diagonal neighbors absent from GK12, which matter most when the local
relation is coupled with AKRR.

In the present calculation, GK16 evaluates \(Q_{\alpha}^{\rm GK}(N,Z)\) from
the 24 surrounding nuclei: experimental \(Q_\alpha\) values whenever
available, AKRR predictions built upon WS4 otherwise. The resulting
\(Q_{\alpha}^{\rm GK}(N,Z)\) replaces the raw AKRR value at \((N,Z)\). This
composite prescription, AKRR+GK16, is the second arrow of
Eq.~\eqref{eq:pipeline}. Note that the replacement acts only on $Q_\alpha$:
GK16 does not feed back into the mass values, so each nucleus enters the
recursion of Eq.~\eqref{eq:mass_recursion} exactly once, as a corrected
link.

\section{Results and Discussion}
\label{sec:results_discussion}

\subsection{Validation on nuclei measured only after AME2003}
\label{sec:test}
\begin{figure}[!htbp]
\includegraphics[angle=0,width=1\textwidth]{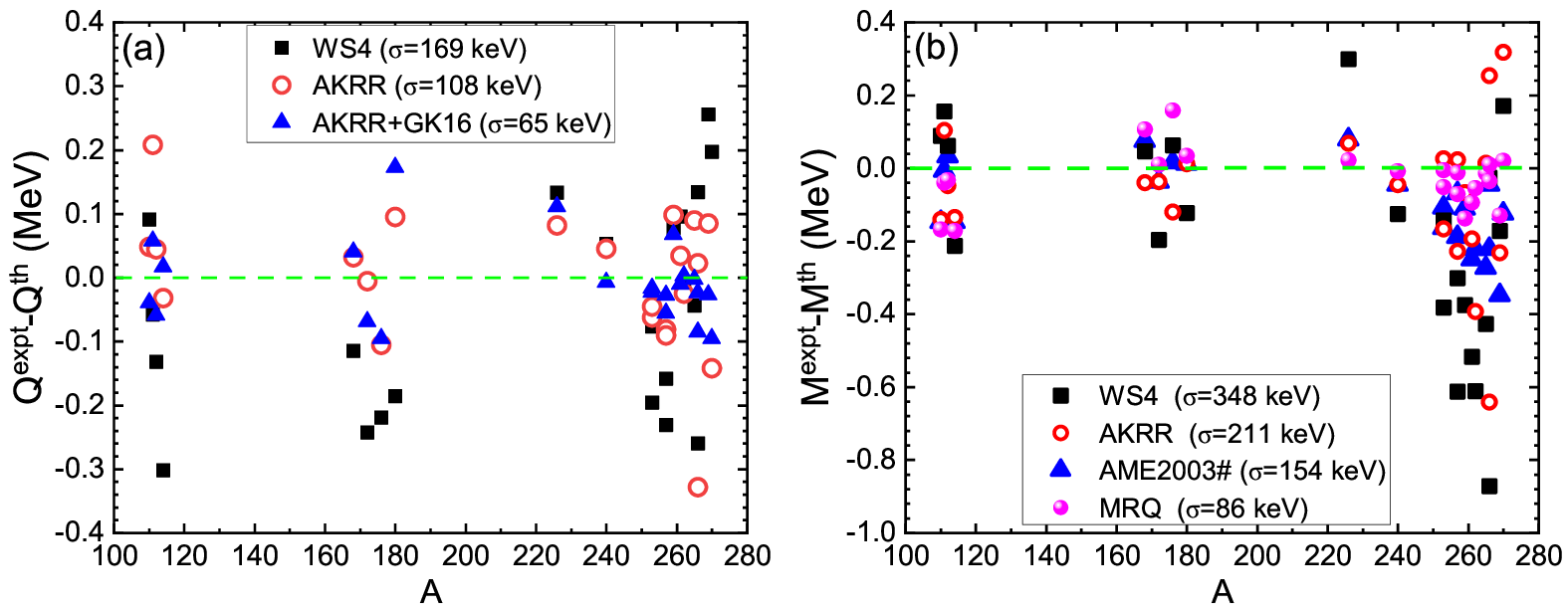}
\caption{(Color online) (a) Differences between the AME2020 experimental and calculated
$\alpha$-decay energies for the 22 benchmark nuclei. The solid squares, open
circles, and solid triangles denote WS4~\cite{Wangn14}, AKRR~\cite{Wuxh24}, and
AKRR+GK16, respectively. (b) Differences between the AME2020 experimental mass
excesses~\cite{AME2020} and the AME2003$\#$ estimates~\cite{Audi03}, WS4, AKRR,
and MRQ results. The RMSD $\sigma$ is given in the legend.}
\label{fig2}
\end{figure}

Following the validation strategy of previous AKRR-based mass
studies~\cite{Tian25}, we benchmark the 22 nuclei with
$110\leq A\leq270$ whose masses are measured in AME2020~\cite{AME2020}
but only estimated in AME2003~\cite{Audi03} (denoted AME2003$\#$
below); all have measured $Q_\alpha$ values in AME2020. The two mass evaluations, taken  nearly two decades apart, thus provide a clean out-of-sample test. To rule out
information leakage, the AKRR network is trained exclusively on the
experimental $Q_\alpha$ values of the 1813 nuclei with uncertainties below 100 keV in
AME2003, i.e., on information predating AME2020. The three hyperparameters were also re-optimized using only the AME2003 
$Q_\alpha$ set; the optimum remained $\lambda=0.3$, $\sigma_1=3.3$, $\sigma_2=0.7$, so the benchmark does not rely on hyperparameters tuned with post-AME2003 data. The AME2020
$Q_\alpha$ values of these 22 nuclei enter only the mass recursion of
Eq.~\eqref{eq:mass_recursion}, as the target's own link or an
intermediate link of another benchmark chain---exactly as in
Sec.~\ref{sec:mass122} for the 122 target nuclei. They are adopted
whenever experimental values exist, irrespective of the 100~keV
uncertainty cut, and never enter the AKRR training set nor serve as
GK16 neighbors. GK16 neighbors of missing links are drawn from
AME2003 with uncertainties below 100~keV; a missing link among the 22
nuclei uses its AME2003 value (or the corresponding AKRR prediction),
and all remaining neighbors are AKRR predictions. The uncertainty of
each adopted $Q_\alpha$ link is propagated to the final mass
uncertainty. The predicted $Q_\alpha$ values and mass excesses are
compared with WS4~\cite{Wangn14} and the standard
AKRR~\cite{Wuxh24} against the AME2020 experimental values in
Fig.~\ref{fig2}.

\begin{table}[!htbp]
\centering
\footnotesize
\caption{\label{tab:benchmark} Experimental and calculated $\alpha$-decay
energies and mass excesses for the 22 benchmark nuclei (in MeV). The
experimental $Q_\alpha$ values and the AME2020 mass excesses are compared with
the WS4, AKRR (denoted AK), and AKRR+GK16 (denoted AK+GK) $Q_\alpha$
predictions, and with the AME2003$\#$, WS4, AK, and MRQ mass excesses,
respectively.}
\begin{ruledtabular}
\begin{tabular}{cccccccccc}
Nuclide & $Q_{\alpha}^{\rm expt}$ & $Q_{\alpha}^{\rm WS4}$ &
$Q_{\alpha}^{\rm AK}$ & $Q_{\alpha}^{\rm AK+GK}$ & $M_{\rm AME2020}$ &
$M_{\rm AME2003\#}$ & $M_{\rm WS4}$ & $M_{\rm AK}$ & $M_{\rm MRQ}$\\
\hline
$^{110}$I	  &	3.581	  &	3.489	  &	3.532	  &	3.619	  &	-60.468	&	-60.321	&	-60.557	&	-60.327	&	-60.296	$\pm$	0.229	\\
$^{111}$I	  &	3.275	  &	3.333	  &	3.066	  &	3.217	  &	-64.954	&	-64.947	&	-65.109	&	-65.057	&	-64.906	$\pm$	0.153	\\
$^{112}$I	  &	2.957	  &	3.089	  &	2.913	  &	3.016	  &	-67.063	&	-67.096	&	-67.125	&	-67.015	&	-67.012	$\pm$	0.147	\\
$^{114}$Cs	&	3.357	  &	3.659	  &	3.389	  &	3.340	  &	-54.686	&	-54.539	&	-54.474	&	-54.551	&	-54.511	$\pm$	0.234	\\
$^{168}$Ir	&	6.381	  &	6.496	  &	6.349	  &	6.341	  &	-18.666	&	-18.741	&	-18.713	&	-18.626	&	-18.773	$\pm$	0.133	\\
$^{172}$Au	&	6.923	  &	7.166	  &	6.929	  &	6.992	  &	-9.318	&	-9.282	&	-9.122	&	-9.281	&	-9.328 $\pm$	0.142	\\
$^{176}$Au	&	6.434	  &	6.652	  &	6.539	  &	6.529	  &	-18.521	&	-18.537	&	-18.584	&	-18.401	&	-18.680	$\pm$	0.077	\\
$^{180}$Tl	&	6.706	  &	6.891	  &	6.610	  &	6.533	  &	-9.390	&	-9.403	&	-9.268	&	-9.403	&	-9.425 $\pm$	0.092	\\
$^{226}$Np	&	8.328	  &	8.194	  &	8.246	  &	8.216	  &	32.817	&	32.738	&	32.518	&	32.749	&	32.795 $\pm$	0.093	\\
$^{240}$Cf	&	7.711	  &	7.658	  &	7.666	  &	7.718	  &	57.989	&	58.034	&	58.114	&	58.034	&	57.997 $\pm$	0.193	\\
$^{253}$No	&	8.415	  &	8.610	  &	8.477	  &	8.431	  &	84.359	&	84.466	&	84.502	&	84.333	&	84.363 $\pm$	0.138	\\
$^{253}$Lr	&	8.918	  &	8.994	  &	8.963	  &	8.940	  &	88.524	&	88.687	&	88.905	&	88.690	&	88.574 $\pm$	0.243	\\
$^{257}$Rf	&	9.083	  &	9.241	  &	9.164	  &	9.110	  &	95.866	&	95.934	&	96.167	&	95.842	&	95.878 $\pm$	0.138	\\
$^{257}$Db	&	9.206	  &	9.437	  &	9.296	  &	9.261	  &	100.154	&	100.342	&	100.767	&	100.382	&	100.224	$\pm$	0.244	\\
$^{259}$Db	&	9.619	  &	9.541	  &	9.520	  &	9.551	  &	101.991	&	102.101	&	102.365	&	102.057	&	102.131	$\pm$	0.212	\\
$^{261}$Bh	&	10.500	&	10.405	&	10.466	&	10.511	&	113.079	&	113.329	&	113.596	&	113.272	&	113.173	$\pm$	0.249	\\
$^{262}$Bh	&	10.320	&	10.328	&	10.345	&	10.317	&	114.252	&	114.473	&	114.864	&	114.645	&	114.305	$\pm$	0.265	\\
$^{265}$Hs	&	10.470	&	10.514	&	10.381	&	10.472	&	120.900	&	121.173	&	121.328	&	120.886	&	120.913	$\pm$	0.190	\\
$^{266}$Hs	&	10.346	&	10.211	&	10.323	&	10.369	&	121.140	&	121.185	&	121.166	&	120.886	&	121.130	$\pm$	0.197	\\
$^{266}$Mt	&	10.996	&	11.255	&	11.324	&	11.081	&	127.673	&	127.893	&	128.544	&	128.314	&	127.707	$\pm$	0.266	\\
$^{269}$Ds	&	11.510	&	11.254	&	11.425	&	11.536	&	134.835	&	135.183	&	135.006	&	135.067	&	134.964	$\pm$	0.204	\\
$^{270}$Ds	&	11.117	&	10.920 	&	11.259	&	11.212	&	134.682	&	134.806	&	134.510	&	134.364	&	134.661	$\pm$	0.203	\\
\end{tabular}
\end{ruledtabular}
\end{table}

The accuracy is quantified by the root-mean-square deviation
\begin{equation}
\sigma(Q_\alpha)=\left[\frac{1}{N_{\rm b}}\sum_{i=1}^{N_{\rm b}}
\bigl(Q_{\alpha,\rm exp}^{(i)}-Q_{\alpha,\rm th}^{(i)}\bigr)^2\right]^{1/2},
\end{equation}
where $N_{\rm b}$ is the number of nuclei entering the comparison. For the 22
nuclei, the $Q_\alpha$ RMSDs are 169~keV for WS4, 108~keV for AKRR, and
65~keV for AKRR+GK16 (Fig.~\ref{fig2}(a)): the local GK16 correction
suppresses the residual fluctuations left by AKRR. For the mass excesses
(Fig.~\ref{fig2}(b) and Table~\ref{tab:benchmark}), the RMSDs are 348~keV for WS4, 211~keV for AKRR, 154~keV
for AME2003$\#$, and 86~keV for the present MRQ calculation. MRQ thus reduces the RMS deviation by about 75\%
relative to WS4 and by about 59\% relative to the direct AKRR mass
prediction, and also improves upon the AME2003 systematic estimates.

This improvement reflects the interpolation/extrapolation asymmetry of
Secs.~\ref{sec:intro} and~\ref{sec:mrq}: applied directly to masses, AKRR
gives 211~keV; applied to $Q_\alpha$ and refined by GK16, 65~keV; the
recursion transfers this accuracy to the mass, giving 86~keV. MRQ does not
treat the unknown mass as an independent model prediction but converts mass
extrapolation into a $Q_\alpha$ interpolation through the exact
energy-conservation relation. The residual 86~keV is dominated by the
anchor-mass uncertainty and the accumulated $Q_\alpha$ uncertainties rather
than by the recursion itself; the gap between 65~keV at the $Q_\alpha$ level
and 86~keV at the mass level reflects the anchor and chain-length dependence
discussed in Sec.~\ref{sec:mrq} and quantified in Sec.~\ref{sec:uncertainty}.

\subsection{Mass predictions for the 122 nuclei without directly measured masses}
\label{sec:mass122}

We apply the MRQ approach to the 122 target nuclei with measured $Q_\alpha$
values but no directly measured masses in AME2020, spanning
$107\leq A\leq294$. For each target, the $\alpha$-decay chain is followed
toward lower mass number until a nucleus with an experimentally known mass is
reached; experimental ground-state $Q_\alpha$ values are adopted whenever
available and missing links are supplied by the WS4-based AKRR+GK16
calculation. The target mass excess then follows from Eq.~\eqref{eq:mass_recursion}.

\begin{figure}[!htbp]
\includegraphics[angle=0,width=0.8\textwidth]{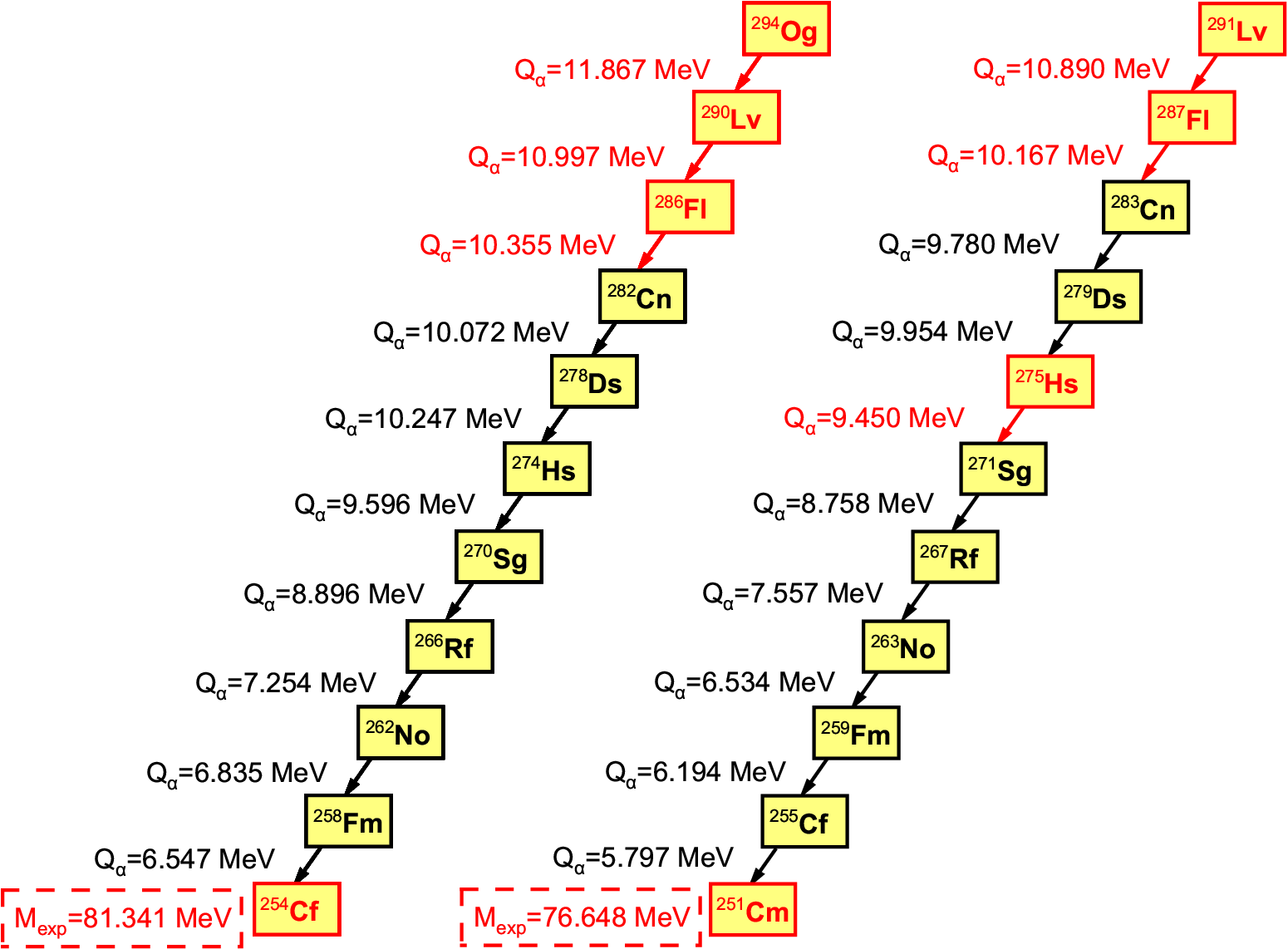}
\caption{(Color online) $\alpha$-decay chains used in the MRQ calculation of
$^{294}_{118}\mathrm{Og}$ and $^{291}_{116}\mathrm{Lv}$. Adopted $Q_\alpha$
values (in MeV) are shown beside the successive links. Red values denote
experimental inputs; black values denote WS4-based AKRR+GK16 predictions used
to complete the chains. The experimental mass excesses of the anchor nuclei
are given at the bottom.}
\label{fig3}
\end{figure}

As a representative example, we take $^{294}_{118}\mathrm{Og}$, the nuclide
with the least direct mass information in Table~\ref{tab:predictions}: its
chain is traced through ten successive $\alpha$ decays, $^{294}$Og $\to$ $^{290}$Lv $\to$
$^{286}$Fl $\to$ $^{282}$Cn $\to$ $^{278}$Ds $\to$ $^{274}$Hs $\to$
$^{270}$Sg $\to$ $^{266}$Rf $\to$ $^{262}$No $\to$ $^{258}$Fm $\to$
$^{254}$Cf, to the experimentally known nucleus
$^{254}_{98}\mathrm{Cf}$. The adopted $Q_\alpha$ values are displayed in
Fig.~\ref{fig3}. Equation~\eqref{eq:mass_recursion} gives
\begin{equation}
M(^{294}_{118}\mathrm{Og}) = M_{\rm exp}(^{254}_{98}\mathrm{Cf})
+\sum_{i=1}^{10}Q_{\alpha i} +10M_\alpha = 198.256~\mathrm{MeV}.
\end{equation}
The same calculation for the analogous chain of $^{291}_{116}\mathrm{Lv}$
yields $M(^{291}_{116}\mathrm{Lv})=185.978~\mathrm{MeV}$. Generally only a
few links require theoretical $Q_\alpha$ values.

The resulting mass excesses for all 122 nuclei are listed in
Table~\ref{tab:predictions}, together with the AME2020 systematic estimates
(AME2020$\#$), available for 112 of them. The AME2020$\#$ values are not
direct measurements; they serve to examine the consistency and extrapolation
behavior of the present predictions, not as an experimental validation set.

\begin{figure}[!htbp]
\includegraphics[angle=0,width=1\textwidth]{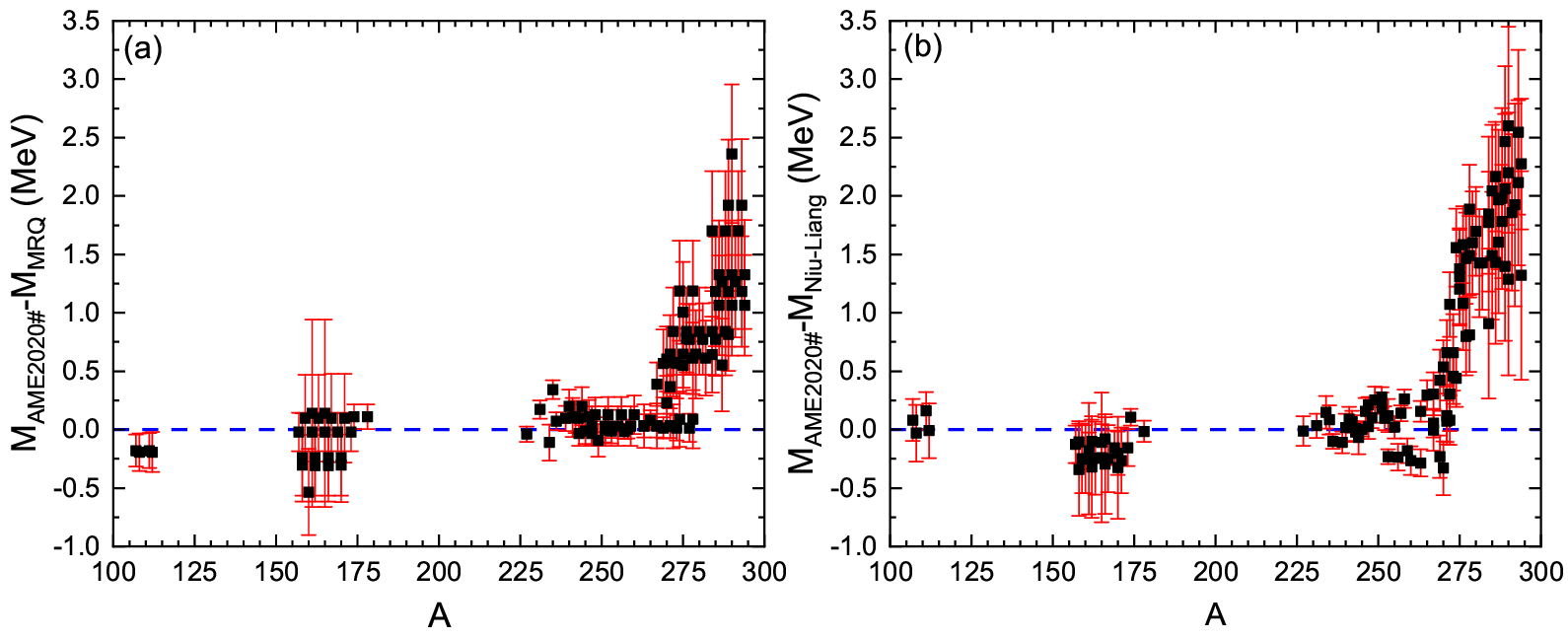}
\caption{(Color online) Mass-excess differences relative to the AME2020$\#$
evaluated values as functions of mass number $A$. (a) AME2020$\#$ minus the
present MRQ predictions. (b) AME2020$\#$ minus the Niu--Liang
results~\cite{Niuzm22}. Both panels use the same vertical scale; dashed
horizontal lines indicate zero difference.}
\label{fig4}
\end{figure}

Figure~\ref{fig4}(a) shows the differences between the AME2020$\#$ estimates
and the present WS4-based MRQ results as functions of mass number. For
$A<265$, the absolute differences remain within about 0.5~MeV, indicating
that the recursion follows the established mass systematics well. Above
$A\approx265$, the differences grow steadily, reaching 1--2~MeV for
$A\ge 288$, with the MRQ masses generally smaller than the AME2020$\#$
values. This behavior is consistent with the cumulative character of the
recursion analyzed in Sec.~\ref{sec:uncertainty}: the farther a target lies
from a measured anchor, the longer the chain and the more missing
$Q_\alpha$ values must be supplied theoretically, so small deviations of
individual decay energies accumulate in Eq.~\eqref{eq:mass_recursion};
moreover, neighboring links come from the same trained network, so their
errors accumulate coherently along the chain. The quoted uncertainties grow
in the same direction: through the position-dependent model term of
Eq.~\eqref{eq:sigmaAKRR}, links farther from the training data carry larger
$\sigma_{Q_\ell}$, and the accumulation transfers this growth to the mass
level. The comparison should nevertheless be interpreted with caution,
since AME2020$\#$ is itself a systematic evaluation rather than a measured
mass; Fig.~\ref{fig4}(b) shows the corresponding differences for the
Niu--Liang predictions~\cite{Niuzm22}, with broadly similar trends. The
growing differences in the superheavy region reflect the divergence between
two extrapolation schemes as both extend away from known nuclei, rather than
the error of either; future mass or $\alpha$-decay-energy measurements in
this region will provide the decisive test.

\subsection{Uncertainty propagation}
\label{sec:uncertainty}

Final-mass uncertainties arise from three sources: experimental errors in training data, AKRR model error that increases away from the training domain, and accumulated uncertainties along decay chains. We perform three-level propagation following the pipeline Eq.~\eqref{eq:pipeline} (AKRR $\to$ GK16 $\to$ recursion), assuming Gaussian, mutually independent residuals.

\emph{(i) AKRR error model and its calibration.}
AKRR predictive uncertainty combines propagated experimental noise and a model-error scale calibrated from LOO residuals~\cite{Fugj11,RW06}.
Define $\boldsymbol{H}=(\boldsymbol{K}+\lambda\boldsymbol{I})^{-1}$ and $\boldsymbol{\beta}=\boldsymbol{H}\boldsymbol{R}_\alpha$. The LOO residual for nucleus $i$, obtained without retraining, is
\begin{equation}
e_i=Q_\alpha^{\rm exp}(i)-Q_\alpha^{\rm AKRR,LOO}(i)
=\frac{[\boldsymbol{H}\boldsymbol{R}_\alpha]_i}{H_{ii}},
\label{eq:eloo}
\end{equation}
where $H_{ii}$ denotes the leverage of nucleus $i$. The experimental noise contribution to the LOO residual reads
\begin{equation}
v_i=\sum_{j\neq i}\Bigl(\frac{H_{ij}\,\sigma_{{\rm exp},j}}{H_{ii}}
\Bigr)^{2};
\label{eq:vin}
\end{equation}
distant nuclei are naturally suppressed by small $H_{ij}$.

The same LOO residuals also calibrate the model-error scale. In region $g$ they are modeled as Gaussian with variance $v_i+\sigma_{{\rm exp},i}^{2}+\tau_g^{2}$, separating known noise from the unknown model deficiency $\tau_g$. We calibrate $\tau_g^{2}$ independently for $A<100$, $100\le A<200$, and $A\ge200$.
Using weights $w_i=1/(v_i+\sigma_{{\rm exp},i}^{2}+\tau_g^{2})$, maximum likelihood yields
\begin{equation}
\mu_g=\frac{\sum_{i\in g}w_ie_i}{\sum_{i\in g}w_i},
\qquad
\tau_g^{2}=\frac{\sum_{i\in g}w_i^{2}\bigl[(e_i-\mu_g)^{2}
-v_i-\sigma_{{\rm exp},i}^{2}\bigr]}{\sum_{i\in g}w_i^{2}},
\label{eq:calib}
\end{equation}
which is solved self-consistently; negative $\tau_g^{2}$ is set to zero when regional scatter is fully explained by propagated noise. The regional bias $\mu_g$ serves only as a diagnostic and is not subtracted. Since $\tau_g^{2}$ is calibrated from the data rather than fitted, no extra free parameter enters at this stage. The regional constant approximation holds within well-sampled training regions; the growth of model error outside these regions is carried into the predictions through the local training-data density.

\emph{(ii) Predictive uncertainty at the AKRR and GK16 levels.}
For nuclide $(N,Z)$, the AKRR prediction is a linear combination of training residuals with coefficients $[\boldsymbol{H}\boldsymbol{k}(N,Z)]_j$, where
$\boldsymbol{k}(N,Z)$ collects kernel values from Eq.~\eqref{eq6}. We introduce a position-dependent weighting factor $\mathcal{W}(N,Z)$ to scale the
calibrated $\tau_g^{2}$:
\begin{equation}
\mathcal{W}(N,Z)=\frac{K[(N,Z),(N,Z)]-\boldsymbol{k}^{\top}(N,Z)\,\boldsymbol{H}\,\boldsymbol{k}(N,Z)}{K[(N,Z),(N,Z)]},
\label{eq:wgp}
\end{equation}
and use its regional average $\langle\mathcal{W}\rangle_g$ to construct the predictive variance
\begin{equation}
\sigma_{\rm AK}^{2}(N,Z)=\tau_g^{2}\,\frac{\mathcal{W}(N,Z)}{\langle\mathcal{W}\rangle_g}
+\sum_{j=1}^{m}\bigl[\boldsymbol{H}\boldsymbol{k}(N,Z)\bigr]_j^{2}\,
\sigma_{{\rm exp},j}^{2}.
\label{eq:sigmaAKRR}
\end{equation}
For the kernel in Eq.~\eqref{eq6}, $K[(N,Z),(N,Z)]=2$ and $0\le\mathcal{W}\le1$: $\mathcal{W}$ is small where data tightly constrain the residual function and approaches unity far from training points, assigning larger model variance to sparsely sampled nuclides. Normalization by $\langle\mathcal{W}\rangle_g$ ties the predictive formula to the LOO calibration. This formulation describes only interpolation uncertainty; it cannot capture new structural features of the WS4 residuals outside the region covered by the training data, such as emerging shell effects. The resulting statistical uncertainties therefore represent lower bounds.

The predictive variance of Eq.~\eqref{eq:sigmaAKRR} then propagates through the GK16 step: experimental $Q_\alpha$ and its uncertainty are adopted when available; otherwise, AKRR predictions with $\sigma_{\rm AK}$ are used. Assuming independent inputs, error propagation through Eq.~\eqref{eq16} gives
\begin{equation}
\sigma_{\rm GK}^2=\frac{1}{16^2}\sum_{j=1}^{24}c_j^2\sigma_j^2,
\label{eq:sigmaGK}
\end{equation}
where $c_j\in\{\pm1,\pm2,6\}$ are the 24 neighbor coefficients of Eq.~\eqref{eq16} and $\sigma_j$ is the uncertainty of neighbor $j$. An incomplete stencil retains the AKRR prediction and its uncertainty. Because neighboring predictions share the WS4 baseline and fitted weights, their errors are positively correlated, making Eq.~\eqref{eq:sigmaGK} a lower bound.

\emph{(iii) Accumulation along the decay chain.}
The uncertainty of the target mass reconstructed recursively via Eq.~\eqref{eq:mass_recursion} is
\begin{equation}
\sigma_{th}^2=\sigma_{\rm anchor}^2+
\sum_{\ell=1}^{n}\sigma_{Q_\ell}^2,
\label{eq:sigmaM}
\end{equation}
where $\sigma_{\rm anchor}$ is the experimental uncertainty of the anchor mass, and $\sigma_{Q_\ell}$ is either the experimental uncertainty of a measured link or $\sigma_{\rm GK}$ for a predicted link; the $^4$He mass-excess uncertainty is negligible. Treating links as independent yields a statistical lower bound: measured links may have negative error correlations leading to partial cancellation, while predicted links inherit the positive correlations discussed above. Again, Sec.~\ref{sec:test} supplies a posteriori validation. Since $\sigma_{Q_\ell}$ inherits the spatial dependence from Eq.~\eqref{eq:sigmaAKRR}, uncertainties for chains extending toward the data frontier grow faster than the simple $\sqrt{n}$ scaling. Equations~\eqref{eq:sigmaAKRR}--\eqref{eq:sigmaM} assign a statistical uncertainty to each predicted mass.

\section{Summary}
\label{sec:summary}
We have presented a nuclear-mass calculation for nuclei without direct mass
measurements that combines AKRR corrections to WS4 $\alpha$-decay energies,
a GK16 local refinement, and recursion from experimentally known anchor
masses. Since all three stages act on $Q_\alpha$, the extrapolation of an
absolute mass is replaced by the interpolation of relative decay energies
followed by exact algebra. The three kernel hyperparameters, fixed by LOO
CV, are the only quantities adjusted to the data, and the propagated
uncertainties are parameter-free as well: a Gaussian-process variance factor
makes the model term grow with the distance to the training data, so the
quoted errors increase along chains that extend toward the data frontier, with
the chain accumulation evaluated at the independent-link lower bound. On the
22 nuclei measured only after AME2003, the $Q_\alpha$ RMSD decreases from
169~keV (WS4) to 108~keV (AKRR) and 65~keV (AKRR+GK16), and the mass-excess
RMSD reaches 86~keV, versus 348, 211, and 154~keV for WS4, AKRR, and the
AME2003$\#$ estimates, respectively. Mass excesses are predicted for the 122
nuclei with measured $Q_\alpha$ values but no directly measured masses;
their differences from the AME2020 systematic estimates and from the
predictions of Niu--Liang~\cite{Niuzm22} grow toward the superheavy region,
where longer chains accumulate the correlated link errors. Future
mass and $\alpha$-decay-energy measurements in this region will tell the
two schemes apart.

\begin{acknowledgments}
\sloppy
This work was supported by the Guangxi Science and Technology Program (No. 2023GXNSFDA026005 and No. 2023GXNSFBA026008), the National Natural Science Foundation of China (No. 12465019 and No. 12465021), and the Central Government Guides Local Scientific and Technological Development Fund Projects (No. Guike ZY22096024).
\end{acknowledgments}

\begingroup
\footnotesize
\setlength{\LTcapwidth}{\textwidth}
\setlength{\tabcolsep}{22pt}
\begin{longtable}{@{}ccccc@{}}
\caption{Predicted mass excesses $M_{\text{MRQ}}$ (in MeV) for the 122
nuclides without directly measured masses, with their experimental
$\alpha$-decay energies $Q_\alpha$ and the AME2020 estimates ($\#$) for
comparison. The column $n$ gives the number of consecutive $\alpha$ decays
in the single decay chain connecting each nuclide to its experimentally
known anchor, i.e., the number of terms in Eq.~\eqref{eq:mass_recursion}.
The uncertainties follow from
Eq.~\eqref{eq:sigmaM}; correlations between predicted links
(Sec.~\ref{sec:uncertainty}) raise the true uncertainty for chains with
several predicted links.}\label{tab:predictions}\\
\hline\hline
Nuclide & $Q_{\alpha}$ & $n$ & $M_{\text{MRQ}}$ & $M_{\text{AME2020\#}}$ \\
\hline
\endfirsthead
\hline\hline
Nuclide & $Q_{\alpha}$ & $n$  & $M_{\text{MRQ}}$ & $M_{\text{AME2020\#}}$ \\
\hline
\endhead
\hline
\endfoot
\hline\hline
\endlastfoot
$^{107}$Te	&	4.010	   & 2  &	-60.478	$\pm$	0.137	&	-60.657\\   
$^{108}$I	  &	4.099	   & 2  &	-52.579	$\pm$	0.160	&	-52.771\\   
$^{111}$Xe	&	3.712	   & 3  &	-54.341	$\pm$	0.149	&	-54.520\\   
$^{112}$Cs	&	3.931	   & 3  &	-46.222	$\pm$	0.171	&	-46.415\\   
$^{157}$Hf	&	5.880	   & 2  &	-38.834	$\pm$	0.166	&	-38.855\\   
$^{158}$Ta	&	6.124	   & 3  &	-30.812	$\pm$	0.261	&	-31.118\\   
$^{158}$W	  &	6.613	   & 3  &	-23.454	$\pm$	0.378	&	-23.693\\   
$^{159}$W	  &	6.451	   & 2  &	-25.531	$\pm$	0.373	&	-25.434\\   
$^{161}$W	  &	5.923	   & 3  &	-30.486	$\pm$	0.166	&	-30.507\\   
$^{160}$Re	&	6.698	   & 3  &	-16.343	$\pm$	0.369	&	-16.878\\   
$^{162}$Re	&	6.240	   & 4  &	-22.147	$\pm$	0.261	&	-22.453\\   
$^{160}$Os	&	7.262	   & 10 &	-8.039 $\pm$	0.944	&	---\\       
$^{161}$Os	&	7.069	   & 10 &	-10.336	$\pm$	0.802	&	-10.196\\   
$^{162}$Os	&	6.768	   & 4  &	-14.261	$\pm$	0.378	&	-14.500\\   
$^{163}$Os	&	6.673	   & 3  &	-16.434	$\pm$	0.373	&	-16.336\\   
$^{165}$Os	&	6.336	   & 4  &	-21.726	$\pm$	0.166	&	-21.747\\   
$^{166}$Ir	&	6.722	   & 5  &	-13.000	$\pm$	0.261	&	-13.306\\   
$^{165}$Pt	&	7.453	   & 11 &	-0.458 $\pm$	0.802	&	-0.318\\    
$^{166}$Pt	&	7.292	   & 5  &	-4.544 $\pm$	0.379	&	-4.783\\    
$^{167}$Pt	&	7.158	   & 4  &	-6.851 $\pm$	0.378	&	-6.753\\    
$^{169}$Pt	&	6.858	   & 5  &	-12.443	$\pm$	0.166	&	-12.464\\   
$^{170}$Au	&	7.177	   & 6  &	-3.398 $\pm$	0.261	&	-3.703\\    
$^{174}$Au	&	6.699	   & 2  &	-14.168	$\pm$	0.105	&	-14.058\\   
$^{170}$Hg	&	7.773	   & 6  &	5.654   $\pm$	0.380	&	5.415\\     
$^{171}$Hg	&	7.668	   & 5  &	3.242   $\pm$	0.378	&	3.339\\     
$^{173}$Hg	&	7.378	   & 6  &	-2.640 $\pm$	0.166	&	-2.661\\    
$^{178}$Tl	&	7.020	   & 3  &	-4.723 $\pm$	0.105	&	-4.613\\    
$^{190}$At	&	7.917	   & 1  &	7.196   $\pm$	0.026	&	---\\       
$^{203}$Ac	&	8.382	   & 1  &	17.579 $\pm$	0.021	&	---\\       
$^{204}$Ac	&	8.107	   & 1  &	16.665 $\pm$	0.034	&	---\\       
$^{207}$Th	&	8.328	   & 1  &	19.354 $\pm$	0.023	&	---\\       
$^{210}$Pa	&	8.445	   & 1  &	24.354 $\pm$	0.067	&	---\\       
$^{214}$U	&	8.696 	   & 1  &	25.180  $\pm$	0.026	&	---\\       
$^{227}$Pu	&	8.338	   & 1  &	36.808 $\pm$	0.065	&	36.770\\    
$^{231}$Am	&	7.234	   & 1  &	42.238 $\pm$	0.079	&	42.410\\    
$^{236}$Am	&	6.256	   & 2  &	45.970 $\pm$	0.076	&	46.041\\    
$^{234}$Bk	&	8.099	   & 2  &	53.505 $\pm$	0.153	&	53.395\\    
$^{235}$Bk	&	7.764	   & 2  &	52.427 $\pm$	0.081	&	52.770\\    
$^{239}$Cf	&	7.764	   & 2  &	58.105 $\pm$	0.107	&	58.202\\    
$^{240}$Es	&	8.259	   & 4  &	64.024 $\pm$	0.142	&	64.225\\    
$^{241}$Es	&	8.259	   & 3  &	63.790 $\pm$	0.166	&	63.893\\    
$^{242}$Es	&	8.160	   & 3  &	64.692 $\pm$	0.139	&	64.801\\    
$^{243}$Es	&	8.072	   & 2  &	64.780 $\pm$	0.107	&	64.747\\    
$^{245}$Es	&	7.909	   & 3  &	66.328 $\pm$	0.109	&	66.315\\    
$^{243}$Fm	&	8.689	   & 3  &	69.219 $\pm$	0.119	&	69.316\\    
$^{247}$Fm	&	8.258	   & 2  &	71.641 $\pm$	0.168	&	71.672\\    
$^{244}$Md	&	8.947	   & 5  &	75.396 $\pm$	0.162	&	75.597\\    
$^{246}$Md	&	8.889	   & 4  &	76.006 $\pm$	0.145	&	76.115\\    
$^{247}$Md	&	8.764	   & 3  &	75.969 $\pm$	0.107	&	75.936\\    
$^{248}$Md	&	8.497	   & 5  &	76.821 $\pm$	0.150	&	76.948\\    
$^{253}$Md	&	7.573	   & 2  &	81.187 $\pm$	0.058	&	81.173\\    
$^{249}$No	&	9.210	   & 3  &	81.881 $\pm$	0.140	&	81.787\\    
$^{251}$No	&	8.752	   & 3  &	82.818 $\pm$	0.168	&	82.849\\    
$^{251}$Lr	&	9.438	   & 4  &	87.832 $\pm$	0.109	&	87.830\\    
$^{252}$Lr	&	9.164	   & 6  &	88.410 $\pm$	0.151	&	88.537\\    
$^{257}$Lr	&	9.068	   & 3  &	92.680 $\pm$	0.066	&	92.665\\    
$^{258}$Lr	&	8.905	   & 3  &	94.778 $\pm$	0.080	&	94.782\\    
$^{253}$Rf	&	9.310	   & 4  &	93.616 $\pm$	0.143	&	93.642\\    
$^{255}$Rf	&	9.056	   & 4  &	94.298 $\pm$	0.168	&	94.329\\    
$^{256}$Db	&	9.336	   & 7  &	100.171	$\pm$	0.154	&	100.298\\   
$^{257}$Sg	&	9.447	   & 5  &	105.488	$\pm$	0.159	&	---\\       
$^{259}$Sg	&	9.765	   & 5  &	106.488	$\pm$	0.169	&	106.519\\   
$^{263}$Sg	&	9.403	   & 2  &	110.162	$\pm$	0.100	&	110.195\\   
$^{265}$Sg	&	8.970	   & 1  &	112.713	$\pm$	0.083	&	112.794\\   
$^{267}$Sg	&	8.400	   & 2  &	115.418	$\pm$	0.186	&	115.806\\   
$^{269}$Sg	&	8.550	   & 3  &	119.127	$\pm$	0.292	&	119.692\\   
$^{260}$Bh	&	10.400   & 8  &	112.996	$\pm$	0.164	&	113.123\\   
$^{270}$Bh	&	9.065	   & 3  &	123.625	$\pm$	0.279	&	124.230\\   
$^{271}$Bh	&	9.420	   & 4  &	125.214	$\pm$	0.335	&	125.859\\   
$^{272}$Bh	&	9.303	   & 5  &	127.945	$\pm$	0.372	&	128.787\\   
$^{274}$Bh	&	8.940	   & 5  &	132.574	$\pm$	0.430	&	133.762\\   
$^{263}$Hs	&	10.734   & 6  &	119.647	$\pm$	0.186	&	119.678\\   
$^{267}$Hs	&	10.038   & 3  &	122.625	$\pm$	0.101	&	122.658\\   
$^{269}$Hs	&	9.340	   & 2  &	124.478	$\pm$	0.092	&	124.493\\   
$^{270}$Hs	&	9.070	   & 4  &	124.885	$\pm$	0.232	&	125.112\\   
$^{271}$Hs	&	9.480	   & 3  &	127.322	$\pm$	0.187	&	127.691\\   
$^{272}$Hs	&	9.772	   & 4  &	128.429	$\pm$	0.292	&	129.004\\   
$^{273}$Hs	&	9.650	   & 4  &	131.202	$\pm$	0.299	&	131.767\\   
$^{275}$Hs	&	9.450	   & 6  &	135.487	$\pm$	0.431	&	136.492\\   
$^{275}$Mt	&	10.483   & 5  &	138.121	$\pm$	0.338	&	138.767\\   
$^{276}$Mt	&	10.100   & 6  &	140.470	$\pm$	0.373	&	141.312\\   
$^{278}$Mt	&	9.580	   & 6  &	144.579	$\pm$	0.431	&	145.767\\   
$^{267}$Ds	&	11.777   & 7  &	133.848	$\pm$	0.193	&	133.880\\   
$^{271}$Ds	&	10.870   & 4  &	135.919	$\pm$	0.103	&	135.952\\   
$^{273}$Ds	&	11.090   & 3  &	138.270	$\pm$	0.106	&	138.285\\   
$^{275}$Ds	&	11.370   & 4  &	141.117	$\pm$	0.188	&	141.666\\   
$^{276}$Ds	&	10.904   & 5  &	141.757	$\pm$	0.294	&	142.539\\   
$^{277}$Ds	&	10.692   & 5  &	144.319	$\pm$	0.299	&	145.092\\   
$^{272}$Rg	&	11.197   & 4  &	142.745	$\pm$	0.187	&	142.773\\   
$^{274}$Rg	&	11.478   & 7  &	144.524	$\pm$	0.237	&	144.612\\   
$^{278}$Rg	&	10.846   & 5  &	149.909	$\pm$	0.307	&	150.521\\   
$^{279}$Rg	&	10.530   & 6  &	151.076	$\pm$	0.378	&	151.721\\   
$^{280}$Rg	&	10.149   & 7  &	153.044	$\pm$	0.373	&	153.886\\   
$^{277}$Cn	&	11.622   & 4  &	152.317	$\pm$	0.121	&	152.332\\   
$^{281}$Cn	&	10.418   & 6  &	157.174	$\pm$	0.306	&	157.947\\   
$^{284}$Cn	&	9.460	   & 8  &	160.715	$\pm$	0.513	&	162.415\\   
$^{278}$Nh	&	11.993   & 8  &	158.941	$\pm$	0.250	&	159.030\\   
$^{282}$Nh	&	10.783   & 6  &	163.117	$\pm$	0.321	&	163.729\\   
$^{284}$Nh	&	10.280   & 8  &	165.749	$\pm$	0.375	&	166.591\\   
$^{285}$Nh	&	10.010   & 8  &	166.585	$\pm$	0.468	&	167.768\\   
$^{286}$Nh	&	9.790	   & 8  &	168.630	$\pm$	0.462	&	169.957\\   
$^{284}$Fl	&	10.721   & 7  &	168.137	$\pm$	0.327	&	168.779\\   
$^{285}$Fl	&	10.600   & 7  &	170.159	$\pm$	0.314	&	170.932\\   
$^{286}$Fl	&	10.324   & 8  &	170.542	$\pm$	0.426	&	171.606\\   
$^{287}$Fl	&	10.167   & 9  &	172.663	$\pm$	0.470	&	173.929\\   
$^{288}$Fl	&	10.060   & 9  &	173.216	$\pm$	0.513	&	174.917\\   
$^{289}$Fl	&	9.954	   & 9  &	175.545	$\pm$	0.561	&	177.465\\   
$^{290}$Fl	&	9.856	   & 9  &	176.369	$\pm$	0.592	&	178.731\\   
$^{286}$Mc	&	10.862   & 7  &	176.404	$\pm$	0.322	&	---\\       
$^{287}$Mc	&	10.760   & 8  &	177.195	$\pm$	0.397	&	177.748\\   
$^{288}$Mc	&	10.650   & 9  &	178.824	$\pm$	0.378	&	179.666\\   
$^{289}$Mc	&	10.490   & 9  &	179.500	$\pm$	0.471	&	180.683\\   
$^{290}$Mc	&	10.410   & 9  &	181.465	$\pm$	0.464	&	182.792\\   
$^{288}$Lv	&	11.240   & 8  &	181.802	$\pm$	0.327	&	---\\       
$^{289}$Lv	&	11.057   & 8  &	183.640	$\pm$	0.314	&	184.457\\   
$^{290}$Lv	&	10.997   & 9  &	183.964	$\pm$	0.430	&	185.028\\   
$^{291}$Lv	&	10.890   & 10 &	185.978	$\pm$	0.478	&	187.244\\   
$^{292}$Lv	&	10.791   & 10 &	186.433	$\pm$	0.513	&	188.133\\   
$^{293}$Lv	&	10.677   & 10 &	188.647	$\pm$	0.565	&	190.568\\   
$^{293}$Ts	&	11.320   & 10 &	193.245	$\pm$	0.474	&	194.428\\   
$^{294}$Ts	&	11.180   & 10 &	195.070	$\pm$	0.466	&	196.397\\   
$^{294}$Og	&	11.867   & 10 &	198.256	$\pm$	0.431	&	199.320\\   
\end{longtable}
\endgroup

\end{document}